\documentstyle[12pt,caption]{article}
\newcommand {\ignore}[1]{}

\newcommand{\noi}{\noindent}
\newcommand{\bc}{\begin{center}}
\newcommand{\ec}{\end{center}}

\def\ifmath#1{\relax\ifmmode #1\else $#1$\fi}
\def\3quarter{{\textstyle{3 \over 4}}}

\def\ra{\rightarrow}

\def\lf{\leaders\hbox to 1em{\hss.\hss}\hfill}

\def\21{$SU(2) \ot U(1)$}
\def\321{$SU(3) \ot SU(2) \ot U(1)$}
\def\ne{\hbox{$\nu_e$ }}
\def\nm{\hbox{$\nu_\mu$ }}
\def\nt{\hbox{$\nu_\tau$ }}
\def\ns{\hbox{$\nu_{sterile}$ }}

\def\Nt{\hbox{$N_\tau$ }}
\def\ns{\hbox{$\nu_S$ }}

\def\O{\hbox{$\cal O$ }}

\def\mnt{\hbox{$m_{\nu_\tau}$ }}

        \def\etc{\hbox{\it etc. }}
        
\def\etal{\hbox{\it et al., }}

\def\rh{\hbox{right-handed }}

\def\gau{\hbox{gauge }}

\def\neu{\hbox{neutrino }}
\def\sa{\hbox{such as }}

\def\neus{\hbox{neutrinos }}

\def\neusc{\hbox{neutrinos, }}

\def\eq#1{{eq. (\ref{#1})}}

\def\VEV#1{\left\langle #1\right\rangle}

\def\lsim{\raise0.3ex\hbox{$\;<$\kern-0.75em\raise-1.1ex\hbox{$\sim\;$}}}
\def\gsim{\raise0.3ex\hbox{$\;>$\kern-0.75em\raise-1.1ex\hbox{$\sim\;$}}}

\def\bel{\begin{letter}}
\def\eel{\end{letter}}
\def\beq{\begin{equation}}
\def\eeq{\end{equation}}
\def\bef{\begin{figure}}
\def\eef{\end{figure}}
\def\bet{\begin{table}}
\def\eet{\end{table}}
\def\bea{\begin{eqnarray}}
\def\ba{\begin{array}}
\def\ea{\end{array}}
\def\bi{\begin{itemize}}
\def\ei{\end{itemize}}
\def\ben{\begin{enumerate}}
\def\een{\end{enumerate}}
\def\ra{\rightarrow}
\def\ot{\otimes}

\def\eea{\end{eqnarray}}
\def\apj#1#2#3{          {\it Astrophys. J. }{\bf #1} (19#2) #3}

\def\aa#1#2#3{          {\it Astron. \& Astrophys.  }{\bf #1} (19#2) #3}

\def\jel#1#2#3{         {\it Journal Europhys. Lett. }{\bf #1} (19#2) #3}

\def\ib#1#2#3{           {\it ibid. }{\bf #1} (19#2) #3}
\def\nat#1#2#3{          {\it Nature }{\bf #1} (19#2) #3}
\def\nps#1#2#3{          {\it Nucl. Phys. B (Proc. Suppl.) }
                         {\bf #1} (19#2) #3}
\def\np#1#2#3{           {\it Nucl. Phys. }{\bf #1} (19#2) #3}
\def\pl#1#2#3{           {\it Phys. Lett. }{\bf #1} (19#2) #3}
\def\pr#1#2#3{           {\it Phys. Rev. }{\bf #1} (19#2) #3}
\def\prep#1#2#3{         {\it Phys. Rep. }{\bf #1} (19#2) #3}
\def\prl#1#2#3{          {\it Phys. Rev. Lett. }{\bf #1} (19#2) #3}
\def\pw#1#2#3{          {\it Particle World }{\bf #1} (19#2) #3}

\def\zp#1#2#3{           {\it Zeit. fur Physik }{\bf #1} (19#2) #3}

\def\n.c.#1#2#3{         {\it Nuovo Cim. }{\bf #1} (19#2) #3}
\def\r.n.c.#1#2#3{       {\it Riv. del Nuovo Cim. }{\bf #1} (19#2) #3}
\def\sjnp#1#2#3{         {\it Sov. J. Nucl. Phys. }{\bf #1} (19#2) #3}

\def\zetfpr#1#2#3{         {\it Z. Eksp. Teor. Fiz. Pisma. Red. }{\bf #1}
(19#2) #3}

\def\mpl#1#2#3{          {\it Mod. Phys. Lett. }{\bf #1} (19#2) #3}

\def\ppnp#1#2#3{           {\it Prog. Part. Nucl. Phys. }{\bf #1} (19#2) #3}

\def\tp{these proceedings}

\def\opc{\hbox{{\sl op. cit.} }}

\relax
\begin{document}
\begin{center}

\newcommand{\dis}{\displaystyle}

{\Large \bf NEUTRINO-LESS DOUBLE BETA DECAY AND BEYOND
THE STANDARD MODEL PHYSICS }
\vskip 1cm
{\large {\bf Jos\'e W. F. Valle \\}
E-mail valle flamenco.ific.uv.es}

\vskip 1cm

{\sl Instituto de F\'{\i}sica Corpuscular - C.S.I.C.,
Departament de F\'{\i}sica Te\`orica,\\
Universitat de Val\`encia, 46100 Burjassot, Val\`encia, SPAIN}\\

\end{center}

\begin{abstract}

A brief sketch is given of the present observational status
and future prospects of the physics of neutrino mass,
including a survey of the various theoretical schemes
of neutrino mass generation. Emphasis is given to those
which are motivated by present experimental hints from
solar and atmospheric neutrinos, as well as from
cosmological data related to the dark matter question.
The conceptual importance of neutrino-less double beta
decay as a distinctive signature of the Majorana character of
neutrinos is stressed. Barring accidental cancellations this process
gives the strongest laboratory constraint on neutrino mass.

\end{abstract}

\section{Introduction}

One of the most fundamental open issues of present-day
particle physics is the determination and theoretical
understanding of the question of whether neutrinos
behave "trivially", as postulated in the standard model,
or whether they have new properties. These include a
wide variety of phenomena such as non-vanishing
mass, oscillations, new interactions, neutrino
decay modes and other exotic properties.

{}From the theoretical point of view, neutrinos are
the only fermions in the standard model without \rh
partners. Although they are also the only electrically
neutral fermions in the theory, lepton number cannot be
broken by a Majorana mass term, because the electroweak
breaking sector is kept to a minimum one consisting of
just one Higgs doublet.
While this cannot be excluded, it is rather mysterious
why neutrinos are so special when compared with the other
fundamental fermions. That this exceptional character of
neutrinos is a circumstantial property of the standard
model is better appreciated when one tries to extend
the theory. Indeed, many unified extensions of the
standard model, such as SO(10), do require the presence
of \rh neutrinos in order to realize the larger symmetry,
and also some extra Higgs representations. These
are needed to break the extra symmetry and may
naturally give rise to  \neu Majorana mass terms.
Moreover, they provide a natural mechanism,
called {\sl seesaw}, to understand the relative smallness
of \neu masses \cite{GRS,fae,nubooks}. In some of these
extensions one can relate the V-A nature of the weak
interaction to the observed smallness of the neutrino
mass, and incorporate parity as a spontaneously
broken symmetry. Unfortunately, the seesaw
mechanism is just a general scheme which cannot,
by itself, provide detailed predictions for
neutrino masses and mixings. These will depend, among other
factors, upon the structure of the Dirac-type
entries, and also on the possible texture of the large
Majorana mass term \cite{Smirnov}.

Although attractive, the seesaw mechanism is by no means
the only way to generate \neu masses. There are  other
attractive possibilities, some of which do not require the
existence of any new gauge symmetries at a large mass scale.
The extra particles required to generate the \neu masses can
have masses at scales accessible to present experiments
\cite{zee.Babu88}. Lepton number (or B-L),
instead of being part of the gauge symmetry \cite{LR}
is simply a global symmetry which is either broken explicitly
or may be spontaneously broken \cite{CMP}. The scale
at which such a symmetry gets broken can be rather
low, close to the weak scale \cite{JoshipuraValle92,MASI_pot3,RPMSW}.
Such a low scale for lepton number breaking could have
important implications not only in astrophysics and
cosmology but also in particle physics.

This large diversity of possible schemes and the
lack of a fundamental theory for the Yukawa couplings imply
that present theory is not capable of predicting the
scale of neutrino masses any better than it can fix the masses
of the other fermions, like that of the muon. As a
result one should at this point turn to experiment.

\subsection{Laboratory Limits on Neutrino Mass}

There are several limits on \neu masses that follow
from observation. The laboratory bounds may be
summarised as \cite{PDG94}
\beq
\label{1}
m_{\nu_e} 	\lsim 5 \: \rm{eV}, \:\:\:\:\:
m_{\nu_\mu}	\lsim 250 \: \rm{keV}, \:\:\:\:\:
m_{\nu_\tau}	\lsim 23  \: \rm{MeV}
\eeq
and follow purely from kinematics. These are the
most model-independent of the \neu mass limits.
The improved limit on the \ne mass from beta decays
was recently given by Lobashev \cite{Erice}, while that
on the \nt mass follows from recent ALEPH data \cite{eps95}
on tau decays to five pions. A  future tau factory should
have enough sensitivity in order to substantially improve
this \nt mass limit \cite{jj}.

In addition, there are limits on neutrino masses
that follow from the non-observation of neutrino
oscillations \cite{granadaosc}. They involve
\neu mass differences versus mixing, and disappear
in the limit of unmixed neutrinos. The present
situation as well as future prospects to probe
for neutrino oscillation parameters at long baseline
experiments is given in Figure \ref{long}.
\bef
\vspace{8.4cm}
\caption{
Oscillation parameters probed at present and future
neutrino experiments}
\label{long}
\eef

\subsection{Neutrino-less Double Beta Decay}

The most stringent limit on neutrino mass arises
from the non-observation of ${\beta \beta}_{0\nu}$
decay, i.e. the process by which nucleus $(A,Z-2)$ decays to
$(A,Z) + 2 \ e^-$. This lepton number violating
process would arise from Majorana \neu exchange.
Although highly favoured by phase space
over the usual $2\nu$ mode, the neutrino-less
process proceeds only if the virtual neutrino
is a Majorana particle. Its decay amplitude is
proportional to a weighted average \neu mass parameter
\beq
\VEV{m} = \sum_{\alpha} {K_{e \alpha}}^2 m_{\alpha}
\label{AVERAGE}
\eeq
where the sum over $\alpha$ includes all light neutrinos.
The negative searches for ${\beta \beta}_{0\nu}$
in $^{76} \rm{Ge}$ and other nuclei leads to the limit  \cite{Avignone}
\beq
\label{bb}
\VEV{m} \lsim 1 - 2 \ eV
\eeq
depending somewhat on the nuclear matrix elements,
characterising this process. Better sensitivity should
be reached at the enriched germanium experiments.
Although the limit in \eq{bb} is rather stringent,
the parameter $\VEV{m}$ in \eq{AVERAGE} can be very
small even though
the neutrino masses themselves are large. This will
happen if there are strong cancellations between the
contributions of different neutrino species. This is
expected to happen accidentally or by virtue of some
symmetry. For example, it happens in the case of a
Dirac \neu because the lepton number symmetry \cite{QDN}
implies the automatic vanishing of $\VEV{m}$.
Even if all \neus are Majorana particles, the parameter
$\VEV{m}$ may differ substantially from the true neutrino
masses $m_\alpha$ relevant for kinematical studies, which
makes these studies therefore complementary.

The ${\beta \beta}_{0\nu}$ decay process may
also be engendered through the exchange of scalar
bosons, thus raising the question of which relationship the
${\beta \beta}_{0\nu}$ decay process bears
with the Majorana nature of the neutrino mass.
A simple but essentially rigorous proof
was given \cite{BOX} to show that, in a gauge theory
of the weak interactions, a non-vanishing
${\beta \beta}_{0\nu}$ decay rate requires \neus to be
Majorana particles, irrespective of which mechanism
induces the process. This old argument (see fig. \ref{box})
relies on the fact that any Feynman graph inducing
neutrino-less double beta decay can be closed, by W exchange,
so as to produce a diagram generating a nonzero Majorana
neutrino mass.
\bef
\vspace{4cm}
\caption{${\beta \beta}_{0\nu}$ decay and Majorana neutrinos.}
\label{box}
\eef
 This establishes a very deep connection
between the two. Unfortunately, only in some special models,
this may be translated into a useful lower limit on the \neu
masses. This happens, for example, if it is induced
mainly by V-A or V+A currents, like in left-right
symmetric models.

\subsection{Hints for Neutrino Mass}

In addition to the above limits there are
some positive {\sl hints} for neutrino masses
that follow from the following
astrophysical and laboratory observations.

The data collected up to now by Homestake and Kamiokande,
as well as by the low-energy data on pp neutrinos from
the GALLEX and SAGE experiments still pose a persisting
puzzle \cite{Davis,granadasol}.
Comparing the data of GALLEX with the Kamiokande data
indicates the need for a reduction of the $^7 $ Be flux
relative to the standard solar model expectation.
The situation can be well represented by a plot
in the plane defined by the $^7 $ Be and $^8$ Be
fluxes \cite{fiorentini}. The one sigma region
for these fluxes allowed by Kamioka and GALLEX data
is obtained as the intersection of the
region to the left of line labelled 91 with the region
labelled KAMIOKA in Figure \ref{solardata}.
The lines are normalised with respect to the reference
solar model of Bahcall and collaborators, but the
argument is model independent. If we include
the Homestake data would of course only aggravate the
discrepancy, since it would, at face value, require a
negative  $^7 $ Be flux, as can be seen from Figure
\ref{solardata}! This strongly suggests that
the solar \neu problem is indeed a real problem,
\bef
\vspace{9.1cm}
\caption{
Allowed one sigma bands for $^7 $ Be and $^8$ Be fluxes
from all solar neutrino data}
\label{solardata}
\eef
and that the simplest astrophysical
solutions to the solar \neu data are disfavoured,
if not ruled out. This therefore suggests
that one needs new physics in the \neu
sector if one wishes to account for the totality of solar
neutrino data \cite{NEEDNEWPHYSICS}.
The most attractive possibility is to assume the
existence of \neu conversions involving very small
\neu masses around $10^{-3}$ eV, as seen in
Figure 4 \cite{MSW}.
The region of parameters allowed by present
experiments is given in ref. \cite{Hata.MSWPLOT}.
Note that the fits favour the non-adiabatic over the
large mixing solution, due mostly to the larger reduction
of the $^7 $ Be flux found in the former.
\bef
\vspace{9.1cm}
\caption{Region of solar \neu oscillation parameters
allowed by experiment}
\label{msw}
\eef

Another possible solution of the solar neutrino
problem  is provided by long wavelength or just-so
neutrino oscillations \cite{rossi95}.


An apparent decrease in the expected flux of atmospheric
$\nu_\mu$'s relative to $\nu_e$'s arising from the decays
of $\pi$'s, $K$'s and secondary muon decays produced in
the atmosphere, has been observed in two underground
experiments, Kamiokande and IMB, and possibly also at
Soudan2 \cite{atm}. Although the predicted absolute
fluxes of \neus produced by cosmic-ray interactions in the
atmosphere are uncertain at the 20 \% level, their
ratios are expected to be accurate to within 5 \%.
This atmospheric neutrino deficit can be ascribed to
\neu oscillations.
Combining these experimental results with observations
of upward going muons made by Kamiokande, IMB and Baksan,
and with the negative Frejus and NUSEX results \cite{up}
leads to the following range of neutrino oscillation
parameters $\Delta m^2_{\mu \tau} \approx 0.005 \: - \: 0.5\ \rm{eV}^2$
and $\sin^22\theta_{\mu \tau} \approx 0.5$.
Similar analyses can be made for the case of \nm to \ns
and \nm to \ne channels, where matter effects play a role
\cite{lipari}.
Recent results from Kamiokande on higher energy \neus
strengthen the case for an atmospheric \neu problem
\cite{atm1} as shown in Figure \ref{kamglasgow}.
\bef
\vspace{9.1cm}
\caption{Region of atmospheric \neu oscillation parameters
from Kamiokande data.}
\label{kamglasgow}
\eef
These data prefer maximum mixing between \nm and \nt
as suggested by theoretical models \cite{DARK92B}.

\subsection{Cosmology and Neutrino Mass}

In addition to laboratory limits, there is a cosmological
bound that follows from avoiding the overabundance of
relic neutrinos \cite{KT}
\beq
\sum_i m_{\nu_i} \lsim 95  \Omega h^2 \: \rm{eV}
\label{rho1}
\eeq
This limit only holds if \neus are stable on cosmological
time scales. There are many models where neutrinos decay
into a lighter \neu plus a Majoron \cite{fae},
\beq
\nu_\tau \ra \nu_\mu + J \:\: .
\label{NUJ}
\eeq
Lifetime estimates in various Majoron models have
been discussed in ref. \cite{V}. These decays can
be fast enough to obey the cosmological limits coming
from the critical density requirement, as well as those
that come from primordial big-bang nucleosynthesis
\cite{BBNUTAU}. Note also that, since these decays
are $invisible$, they are consistent with all
astrophysical observations.
\bef
\vspace{8.4cm}
\caption{
Estimated \nt lifetime versus observational limits.
}
\label{ntdecay}
\eef

Relic neutrinos may also disappear by annihilation
into Majorons,
\beq
\nu_\tau  \nu_\tau \ra J J \: .
\label{nunujj}
\eeq
thus easily making it possible to obey
cosmological nucleosynthesis limits  \cite{drpv}.


Recent observations of cosmic background temperature
anisotropies on large scales by the COBE  satellite
\cite{cobe} seem to be in conflict with the simple
cold dark matter model of structure formation, if
one adopts as normalisation of the density fluctuation
the cluster-cluster correlation data obtained e.g.
from IRAS \cite{iras}. These data indicate the need for
having more power on large scale. One of the simplest
ways to achieve this is to postulate the existence
of a hot {\sl dark matter} component, contributing
about 30\% to the total mass density \cite{cobe2}.
A good fit is provided by a massive neutrino, for
example, a tau neutrino with mass in the few eV range.
If such is the case one may expect the possibility
of observing \ne to \nt or \nm to \nt oscillations
in the CHORUS and NOMAD experiments at CERN, as well
as at the at the proposed P803 experiment at Fermilab
\cite{chorus}. This mass scale is also consistent with
the recent preliminary indications that might favour
the existence of neutrino oscillations from the LSND
experiment \cite{Caldwell}.


Al alternative way to reconcile COBE and IRAS
observations is the idea that a late decaying
tau neutrino with mass in the MeV range can delay
the epoch of radiation matter equality and will
be discussed below \cite{ma1}.

\section{Reconciling Present Hints.}

Reconciling the present hints from astrophysics and
cosmology in the framework of a consistent elementary
particle physics model is not straightforward.
It requires the existence of a neutrino
with a mass scale which is clearly at odds
with those inferred from the solar and atmospheric neutrino
data discussed above. Indeed, if all the data are taken at
face value they put an interesting theoretical puzzle whose
possible resolutions will now be discussed.

\subsection{Three Almost Degenerate Neutrinos}
\label{Degenerate}

The only possibility to reconcile the above three
 observations in a world with just the three neutrinos
of the standard model is if all of them have nearly the
same mass $\sim$ 2 eV \cite{caldwell}. This is clearly
at odds with the simplest seesaw model where the neutrino
masses scale as those of the up-type quarks.
However, it is known that the general seesaw models have
two independent terms giving rise to the light neutrino masses.
The first is proportional to an effective triplet vacuum expectation value
\cite{2227} which is expected to be small in left-right
symmetric models \cite{LR}. Based on this fact one can
in fact construct extended seesaw models where the main
contribution to the light \neu masses ($\sim$ 2 eV) is universal,
due to a suitable horizontal symmetry, while the splittings
between \ne and \nm explain the solar \neu deficit and that
between \nm and \nt explain the atmospheric \neu anomaly \cite{DEG}.

If this all hints for neutrino mass are taken seriously
and one adopts a minimal three neutrino scenario to explain
them in terms of neutrino properties, it follows that the
observability of neutrino-less double beta decay at the
next generation of enriched germanium experiments
should be possible.

\subsection{Three Active plus One Sterile Neutrino}

The alternative way to fit all the data is to add a
fourth \neu species which, from the LEP data on the
invisible Z width, we know must be of the sterile type,
call it \ns. The first scheme of this type gives mass
to only one of the three neutrinos at the tree level,
keeping the other two mass-less \cite{OLD}.
In a seesaw scheme with broken lepton number, radiative
corrections involving gauge boson exchanges will give
small masses to the other two neutrinos \ne and \nm
\cite{Choudhury}. However, since the singlet \neu is
super-heavy in this case, there is no room to account
for the three hints discussed above.

The schemes which have been suggested to keep the sterile
neutrino light require the use of a special symmetry
\cite{DARK92,DARK92B,DARK92C}. Of these models, those
containing only four neutral lepton states,
with the fourth being the sterile \neu \ns, invoke
the existence of additional Higgs bosons beyond that
of the standard model, in order to generate radiatively
the scales required for the solar and atmospheric \neu
conversions. In these models the \ns either lies at the dark matter
scale \cite{DARK92} or, alternatively, at the solar \neu scale
\cite{DARK92B}.
In the first case the atmospheric
\neu puzzle is explained by \nm to \ns oscillations,
while in the second it is explained by \nm to \nt
oscillations. Correspondingly, the deficit of
solar \neus is explained in the first case
by \ne to \nt oscillations, while in the second
it is explained by \ne to \ns oscillations. In both
cases it is possible to fit all observations together.
However, in the first case there is a clash with the
bounds from big-bang nucleosynthesis. In the latter
case the \ns is at the MSW scale so that nucleosynthesis
limits are satisfied. They single out the non-adiabatic
solution uniquely. Note also that the
mixing angle characterising the \nm to \nt
oscillations is nearly maximal, as suggested
from Figure \ref{kamglasgow}, taken from
ref. \cite{atm1}. Moreover, the model would naturally fit the
recent preliminary hints of neutrino oscillations of
the LSND experiment \cite{Caldwell}.
Another theoretical possibility is that all active
\neus are very light, while the sterile \neu \ns is
the single \neu responsible for the dark matter
\cite{DARK92D}.

As a last comment we mention that it has been
argued that schemes where
the hot dark matter is made up of two or more
active neutrinos may provide a better fit of
the data on structure formation than models
with a single such neutrino \cite{shafi95}.

\subsection{KeV Majoron and Late Decaying Tau Neutrino}

Although we regard the late decaying tau neutrino
solution very promising, it has not yet been as well
explored as the mixed cold and hot dark matter picture
of structure formation discussed above, which has been
extensively studied in the last three years \cite{cobe2,shafi95}.
This solution postulates  an MeV scale tau neutrino which
decays with lifetime of order of years. This lifetime value
 would fit nicely the required parameters and, on the
other hand, would be perfectly achievable in elementary
particle physics models where neutrino masses
arise from the spontaneous violation of lepton
number \cite{fae}, as can be seen from fig.
\ref{ntdecay}. There have also been speculations
that the Majoron in these models could  pick up
a mass in the KeV range, as a result of gravitational
effects \cite{Goran92} and that it would be the
present-day cold dark matter particles \cite{KEV}.
This opens the possibility of a complete picture of
cosmological dark matter which does not postulate
two unrelated components of dark matter (like cold
and hot) but successfully fits the observations
from a common physics principle: the massive tau
neutrino acquires mass from the spontaneous violation
of lepton number and the resulting Majoron with mass
in the KeV range plays the role of cold dark matter
\cite{kev95}. In contrast to the usual
collision-less dark matter, these Majorons have a relatively strong
self-interaction, needed in order to comply with nucleosynthesis
constraints on the MeV tau neutrino, in the presence of the
keV Majoron mass. This yields a rather different and
interesting picture of structure formation. Electron and muon
neutrinos would
be very light, as required in order to account for the solar
neutrino deficit through \ne to \nm oscillations \cite{RPMSW}.
Super-symmetry with spontaneously broken R parity \cite{MASI_pot3}
provides a natural particle physics model for this scenario.
Unfortunately this model cannot account for the atmospheric neutrino
anomaly which would require the existence of a fourth sterile neutrino
\cite{JV95}.

\section{New Signatures.}

The important role massive neutrinos can play
in particle physics and cosmology and the existence
of the hints discussed above should encourage one to continue
the efforts to improve present laboratory \neu mass
limits, and/or to search for related signatures.
Indeed, neutrino masses could be responsible for a wide
variety of measurable implications at the laboratory.
These new phenomena would cover an impressive range of energies,
starting with the searches for anomalies in $\beta$ decays \cite{Deutsch},
the searches for nuclear $\beta \beta_{0\nu}$
decays \cite{Avignone}, the searches for neutrino oscillations
at nuclear reactors and accelerators \cite{granadaosc}, especially
those with long baseline, which would help cross checking
the present indications of an atmospheric neutrino anomaly and
so on. It is not so often stressed that the signals related
to neutrino properties beyond the standard model may sometimes
show up even at the highest energies available at present-day particle
colliders. In the following brief and biased survey
of the situation I will give some examples of the latter.

\subsection{Neutrino-less Double Beta Decay}

If neutrinos are massive Majorana particles
one expects neutrino-less double beta decays
to take place at some level. Barring special cancellations
it should be observable at enriched germanium experiments,
if the quasi-degenerate neutrino scenario for the joint
explanation of hot dark matter with the solar and atmospheric
\neu anomalies discussed in section \ref{Degenerate}
is realized in nature.

Gauge theories may also lead to new varieties of
neutrino-less double beta decays, involving
the $emission$ of light scalar bosons, such as
the Majoron, denoted by $J$ \cite{GGN}
and of a related light scalar boson $\rho$
\beq
(A,Z-2) \rightarrow (A,Z) + 2 \ e^- + J \:.
\eeq
The emission of such weakly interacting light
scalars would only be detected through their
effect on the $\beta$ spectrum.
The simplest model leading to sizeable Majoron
emission in $\beta\beta$ decays involving an
isotriplet Majoron \cite{GR} leads to a new
invisible decay mode for the neutral weak interaction \gau
boson with the emission of light scalars,
\beq
Z \ra \rho + J,
\label{RHOJ}
\eeq
now ruled out by LEP measurements of the
invisible Z width \cite{LEP1}.
It has however been shown that a sizeable
Majoron-neutrino coupling leading to observable
emission rates in neutrino-less double beta decay
can be reconciled with the LEP results in models
where the Majoron is an isosinglet and lepton number
is broken at a low scale \cite{ZU}. An alternative
possibility was discussed in ref. \cite{Burgess93}.
At the moment there is only a limit on the Majoron
emitting neutrino-less double beta decay lifetime,
leading to a bound on the Majoron-neutrino coupling
of about $10^{-4}$ \cite{klapdor_wein}.
New varieties of neutrino-less double beta decay
involving multiple $emission$ of light scalars
also exist \cite{Boston} but it is hard to make the associated
rates large enough to be experimentally observable
\cite{bamert95}.

\subsection{Lepton Flavour Violation.}

Another manifestation of neutrino properties
beyond the standard model, such as neutrino masses or
  isosinglet neutral heavy leptons (NHLS),
is the observability of lepton flavour violating (LFV) decays
\sa $\mu \rightarrow e \gamma$. Such decays are exactly forbidden
in the standard model. Although these are a generic
feature of models with massive \neus, they may
proceed in models where \neus are strictly
mass-less \cite{SST,CP,BER}. This is not only of
conceptual importance but also practical. It
means that the expected rate for LFV processes
is necessarily suppressed due to the smallness
of neutrino masses.
Indeed, in these models the relevant constraints come
from the universality of the weak interaction, which
allows for decay branching ratios larger than the present
experimental limits for a wide variety of LFV decays
\cite{3E,Pila}.
The results are summarised in Table 1. As an illustration,
Figure \ref{pila} gives the expectations for the three
charged lepton decays of the tau, taken from ref. \cite{Pila}.
\bef
\vspace{9.1cm}
\caption{Expected branching ratios for $\tau \ra 3e$
(solid) and $\tau \ra \mu \mu e$}
\label{pila}
\eef
Clearly these branching ratios lie within the
sensitivities of the planned tau and B factories,
as shown in ref. \cite{TTTAU}.
\begin{table}
\begin{center}
\caption{Allowed $\tau$ decay branching ratios }.
\begin{displaymath}
\begin{array}{|c|cr|}
\hline
\mbox{channel} & \mbox{strength} & \mbox{} \\
\hline
\tau \rightarrow e \gamma ,\mu \gamma &  \lsim 10^{-6} & \\
\tau \rightarrow e \pi^0 ,\mu \pi^0 &  \lsim 10^{-6} & \\
\tau \rightarrow e \eta^0 ,\mu \eta^0 &  \lsim 10^{-6} - 10^{-7} & \\
\tau \rightarrow 3e , 3 \mu , \mu \mu e, \etc &  \lsim 10^{-6} - 10^{-7} & \\
\hline
\end{array}
\end{displaymath}
\end{center}
\end{table}

The physics of rare $Z$ decays nicely complements what
can be learned from the study of rare LFV muon and tau decays.
The stringent limits on $\mu \rightarrow e \gamma$ preclude any
possible detect-ability at LEP of the corresponding
$Z \rightarrow e \mu$ decay. However the decays with
tau number violation, $Z \ra e\tau$ or $\mu\tau$ can be large,
as shown in Table 2.
Similarly one can show that the CP violating Z decay
asymmetries in these LFV processes can reach \O($10^{-7}$) \cite{CP}.
Under realistic
luminosity and experimental resolution assumptions, however, it is
unlikely that one will be able to see these decays of the Z at
LEP without a high luminosity option \cite{ETAU}.
In any case, there have been dedicated experimental
searches which have set good limits \cite{opalLFV}.

If the NHLS are lighter than the $Z$, they may also be
produced directly in Z decays such as \cite{CERN},
\begin{equation}
Z \rightarrow N_{\tau} + \nu_{\tau}
\end{equation}
Note that the isosinglet neutral heavy lepton
\Nt is singly produced, through the off-diagonal
neutral currents which is characteristic of models containing
doublet and singlet leptons \cite{2227}.
Subsequent \Nt decays would then give rise to
large missing momentum events, called zen-events.
Theoretically attainable rates for such
processes are large (see ref. \cite{CERN}) and the
present limits are summarized in Figure \ref{cern}.
\bef
\vspace{9.1cm}
\caption{Limits on $Z  \ra N \nu$ decays}
\label{cern}
\eef
The LEP limits follow from the negative searches for acoplanar
jets and lepton pairs from $Z$
decays at LEP, although some inconclusive positive hints have
also been reported by the ALEPH collaboration \cite{opalNHL}.
\begin{table}
\begin{center}
\caption{Allowed branching ratios for rare $Z$
decays. }
\begin{displaymath}
\begin{array}{|c|cr|}
\hline
\mbox{channel} & \mbox{strength} & \mbox{} \\
\hline
Z \rightarrow \Nt \nt &  \lsim 10^{-3} & \\
Z \rightarrow e \tau &  \lsim 10^{-6} - 10^{-7} & \\
Z \rightarrow \mu \tau &  \lsim 10^{-7} & \\
\hline
\end{array}
\end{displaymath}
\end{center}
\end{table}

Note also  that there can also be large rates for
lepton flavour violating decays in models with radiative
mass generation \cite{zee.Babu88}. For example, this is
the case in the models proposed to reconcile present
hints for \neu masses \cite{DARK92B,DARK92}. The expected decay
rates may easily lie within the present experimental
sensitivities and the situation should improve at PSI
or at the proposed tau-charm factories.

Finally, another possible type of LFV decays are those
involving the emission of a Majoron, such as
single Majoron emitting $\mu$ and $\tau$
decays \cite{NPBTAU}.
 These would be "seen" as bumps in the final
lepton energy spectrum, at half of the parent lepton
mass in its rest frame. They are present in models
with spontaneous violation of R parity
\cite{MASI_pot3}.
The allowed rates for these decays may fall within
present experimental sensitivities \cite{PDG94}.
As an illustration,
I borrow Figure \ref{npbtau} from ref. \cite{NPBTAU}.
\bef
\vspace{9.1cm}
\caption{Allowed branching ratios for $\tau \ra e + J$ versus \mnt}
\label{npbtau}
\eef
This example also illustrates how the search for
rare decays can be a more sensitive probe of \neu
properties than the more direct searches for \neu
masses, and therefore complementary. Moreover, they are
ideally studied at a tau-charm factory \cite{TTTAU}.

\subsection{Invisibly Decaying Higgs Bosons.}

Nonstandard neutrino properties may affect
even the electroweak breaking sector.
For example, many extensions of the
lepton sector seek to give masses to neutrinos through the
spontaneous violation of an ungauged U(1) lepton number
symmetry, thus implying the existence of a physical
Goldstone boson, called Majoron \cite{CMP}. As already
mentioned above this is consistent with the measurements
of the invisible $Z$ decay width at LEP if the Majoron
is (mostly) a singlet under the \21 \gau symmetry.

Although the original Majoron proposal was made
in the framework of the minimal seesaw model, and
required the introduction of a relatively high
energy scale associated to the mass of the \rh
\neus, there are many attractive
theoretical alternatives where lepton number
is violated spontaneously at the weak scale or
lower. In this case although the Majoron has very
tiny couplings to matter and the \gau bosons, it
can have significant couplings to the Higgs bosons.
The latter can, as a result, decay with a substantial branching ratio
into the invisible mode \cite{JoshipuraValle92}
\begin{equation}
h \rightarrow J\;+\;J
\label{JJ}
\end{equation}
The production and subsequent decay of a Higgs boson
which may decay visibly or invisibly involves three independent
parameters: its mass $M_H$, its coupling strength to the Z,
normalised by that of the standard model, $\epsilon^2$, and its
invisible decay branching ratio.
The LEP searches for various exotic channels can be used
in order to determine the regions in parameter space
that are already ruled out \cite{alfonso}. The result
is shown in Figure \ref{alfonso2}
taken from the first paper in ref. \cite{moriond}.

\bef
\vspace{9.1cm}
\caption{Region in the $\epsilon^2$ vs. $m_H$ that can be
excluded by the present LEP1 analyses (solid curve).
Also shown are the LEP2 extrapolations (dashed).}
\label{alfonso2}
\eef

Another mode of production of invisibly decaying
Higgs bosons is that in which a CP even Higgs boson
is produced at LEP in association with a massive
CP odd scalar. This production mode is
present in all but the simplest Majoron model
and the corresponding LEP limits on the
relevant parameters are given in ref. \cite{HA}.

\ignore{
In this plot we have assumed
BR ($H \rightarrow J\:J$) = 100\% and a visibly
decaying A boson.
\bef
\vspace{9.1cm}
\caption{Limits on $\epsilon^2_{A}$ in the $m_A,m_H$ plane
that can be placed by present LEP1 searches based on the
 $e^+ e^- \rightarrow H \:A \rightarrow J\:J b\bar{b}$
production channel.      }
\label{ha}
\eef
  }
Finally, the invisible decay of the Higgs boson may
also affect the strategies for searches at higher energies.
For example, the ranges of parameters that can be covered
by LEP2 searches for a total integrated luminosity of
500 pb$^{-1}$ and various centre-of-mass energies have
been given in Figure \ref{alfonso2}. Similar analysis were
made for the case of a high energy linear $e^+ e^-$ collider
(NLC) \cite{EE500}, as well as for the LHC \cite{granada}.

\subsection{Conclusion}

\noi
Present cosmological and astrophysical observations,
as well as theory, suggest that neutrinos may be massive.
Existing data do not preclude neutrinos from being responsible
for a wide variety of measurable implications at the laboratory.
It is therefore quite worthwhile to keep
pushing the underground experiments, for any
possible confirmation of \neu masses.
These includes experiments with enriched germanium
looking for neutrino-less $\beta \beta$ decays,
solar \neu experiments GALLEX and SAGE,
as well as Superkamiokande, Borexino, and
Sudbury. The same can be said of the ongoing
studies with atmospheric \neusc which may be
cross-checked at long baseline neutrino
oscillation searches.
Similarly, a new generation of experiments capable
of more accurately measuring the cosmological
temperature anisotropies at smaller angular scales than
COBE, would be good probes of different models of
structure formation, and presumably shed further
light on the possible role of neutrinos as dark matter.


\section*{Acknowledgements}

This work has been supported by DGICYT under
Grant number PB92-0084. I thank Sasha Dolgov for discussions

\bibliographystyle{ansrt}

\end{document}